\documentclass[twocolumn,trackchanges]{aastex62}
\usepackage{natbib}
\usepackage{multirow}
\usepackage{graphicx}
\usepackage{tabularx}
\usepackage{verbatim}
\usepackage{color}
\usepackage{ulem}
\usepackage{subfigure}

\graphicspath{{./}{figures/}}

\submitjournal{ApJ}

\shorttitle{The Gas Accretion History of Low Mass Halos}
\shortauthors{Chun et al.}

\begin{document}

\title{The Gas Accretion History of Low Mass Halos within the Cosmic Web from Cosmological Simulations}

\correspondingauthor{Sungsoo S. Kim}
\email{kwchun@khu.ac.kr, sungsoo.kim@khu.ac.kr}

\author{Kyungwon Chun}
\affil{School of Space Research, Kyung Hee University, 1732 Deogyeong-daero, Yongin-si, Gyeonggi-do 17104, Korea}

\author{Rory Smith}
\affiliation{Korea Astronomy and Space Science Institute (KASI), 776 Daedeokdae-ro, Yuseong-gu, Daejeon 34055, Korea}

\author{Jihye Shin}
\affiliation{Korea Astronomy and Space Science Institute (KASI), 776 Daedeokdae-ro, Yuseong-gu, Daejeon 34055, Korea}

\author{Sungsoo S. Kim}
\affiliation{School of Space Research, Kyung Hee University, 1732 Deogyeong-daero, Yongin-si, Gyeonggi-do 17104, Korea}
\affiliation{Department of Astronomy \& Space Science, Kyung Hee University, 1732 Deogyeong-daero, Yongin-si, Gyeonggi-do 17104, Korea}

\author{Mojtaba Raouf}
\affiliation{Korea Astronomy and Space Science Institute (KASI), 776 Daedeokdae-ro, Yuseong-gu, Daejeon 34055, Korea}

\begin{abstract}
Using high resolution hydrodynamical cosmological simulations, we study the gas accretion history of low mass halos located in a field-like, low density environment. We track their evolution individually from the early, pre-reionization era, through reionization, and beyond until $z=0$. Before reionization, low mass halos accrete cool cosmic web gas at a very rapid rate, often reaching the highest gas mass they will ever have. But when reionization occurs, we see that almost all halos lose significant quantities of their gas content, although some respond less quickly than others. We find that the response rate is influenced by halo mass first, and secondarily by their internal gas density at the epoch of reionization. Reionization also fully ionises the cosmic web gas by z$\sim$6. As a result, the lowest mass halos (M$\sim$10$^6~h^{-1}$~M$_\odot$ at $z=6$) can never again re-accrete gas from the cosmic web, and by $z\sim5$ have lost all their internal gas to ionisation, resulting in a halt in star formation at this epoch. However, more massive halos can recover from their gas mass loss, and re-accrete ionised cosmic web gas. We find the efficiency of this re-accretion is a function of halo mass first, followed by local surrounding gas density. Halos that are closer to the cosmic web structure can accrete denser gas more rapidly. We find that our lower mass halos have a sweet spot for rapid, dense gas accretion at distances of roughly 1-5 virial radii from the most massive halos in our sample ($>$10$^8~h^{-1}$~M$_\odot$), as these tend to be embedded deeply within the cosmic web.
\end{abstract}


\section{Introduction}
Historically, the environment that galaxies inhabit was broadly split into three main categories; at the lowest environmental densities were the relatively isolated field galaxies, at medium densities were groups of galaxies consisting of tens or hundreds of galaxies, like our own Galaxy, and at high densities were clusters of galaxies consisting of thousands of galaxies in a host with mass greater than $\sim$10$^{14}$~M$_\odot$. However, it has subsequently been revealed by large volume redshift surveys \citep{Colless2001,Tegmark2004} that the density of environment that galaxies inhabit is much broader and less easily subcategorized than was previously thought. In fact, galaxies are found in a vast, complex structure known as the `Cosmic-web' \citep{Bond1996}. In some parts of the cosmic web, there are near empty regions known as voids. Voids are surrounded by sheet-like structures that, in turn, are framed by filaments. The traditional galaxy groups tend to be found aligned along the filaments, and where multiple filaments, galaxy clusters are found. Numerical simulations in a Lambda cold dark matter ($\Lambda$CDM) Universe can well reproduce the broad properties of this web-like structure. The dark matter distribution defines the skeleton of the structure, and warm ionised gas is trapped by the potential well of the dark matter in the structure, and tends to flow and shock in the filaments \citep{Codis2012,Laigle2015,Hahn2015}. This cosmic web gas is a crucial source of external fuel for star-formation in galaxies that can accrete from it. Observations of our own galaxy suggest that, without an external source of gas such as the cosmic web, there would be insufficient gas reservoirs to maintain star formation for more than a few gigayears, and yet our galaxy shows evidence for star formation on much longer timescales \citep{Larson1972}.

The accretion of gas from filaments in the cosmic web and its impact on galaxy evolution has been investigated by many previous studies \citep{Keres2005,Keres2009,Kleiner2017,Kraljic2018,Liao2019}.
\cite{Liao2019} suggest that the halos on the filament have higher baryon and stellar fraction compared to their counterparts in the field. \cite{Keres2005,Keres2009} showed that galaxies acquire most of their gas mass through filamentary `cold mode' gas accretion. In particular, they find that cold accretion dominates the mass growth of galaxies in low mass halos at all times.

Galaxies inhabiting low mass halos are generally known as `dwarf galaxies'. They are easily the most common type of galaxies in the Universe. In the standard cosmology model, $\Lambda$CDM, dwarf galaxies have a special importance. They are considered the crucial low mass building blocks of all more massive galaxies, which grow hierarchically through a series of mergers with lower mass halos. In this framework, understanding dwarf galaxy evolution is crucial, and required for a full understanding of the formation and evolution of all galaxies.

As dwarf galaxies inhabit low mass halos, their shallow potential wells are expected to make them highly sensitive to external physical mechanisms, such as environmental effects. Cosmic reionization is believed to have a considerable impact on the fate of the dwarf galaxies.
Following the epoch of reionization, the harsh ultra-violet (UV) radiation is expected to completely ionise the gas contained within the lowest mass halos \citep{Gnedin2000,Hoeft2006,Okamoto2008}. This can effectively halt their star formation, by ionising their fuel for star formation. Low mass halos form preferentially early in the over dense regions where galaxy clusters will eventually form and it has been theorised that this maybe one of the reasons why galaxy clusters are filled with so many old, and passive dwarf galaxies \citep{Tully2002}. 

If reionization can ionise the dense gas contained at the centre of low mass halos, it is natural to expect it to completely ionise the lower density surrounding cosmic web gas from which the low mass halos were previously accreting. Once this gas is ionised it may simply be too hot to be bound to the shallow potential wells of the low mass halos \citep{Gnedin2000,Hoeft2006,Okamoto2008}, and the halos cease to accrete any new gas in the future. 
Recent observational and numerical studies show that the star formation of some dwarf galaxies is almost quenched by cosmic reionization or supernova feedback \citep{Ricotti2005,Bovill2011a,Bovill2011b,Brown2014,Bettinelli2018}  Nevertheless, some dwarf galaxies show extended star formation histories beyond the epoch of reionization, perhaps indicating they can maintain some of their gas against reionization, or later accrete some fresh gas to re-initiate star formation \citep{Ricotti2009,Skillman2014,Weisz2014,McQuinn2017,Wright2019}.

In this paper, we aim to improve our understanding of how dwarf galaxies, whose virial mass $M_{vir}$ is smaller than $10^{10}~h^{-1}M_{\odot}$ at $z=0$, both accrete cosmic web gas, and lose their gas to reionization. We attempt to understand what parameters dictate the efficiency by which they accrete gas, and they convert the accreted gas into dense internal gas that can be used for star formation. For this, we study the gas accretion histories of dark matter halos within three different cosmological hydrodynamic zoom-in simulations of a 1~$h^{-1}$Mpc volume at $z=0$ \citep{Chun2019}.
Given the small box size of our simulations, it is inevitable that there will be some impact from cosmic variance on our results. Nevertheless, as our sample of halos are much smaller than the box size, we believe that our primary result -- assessing the parameter dependence of their gas accretion -- will remain robust to cosmic variance. We confirm that we see little variation in the parameter dependence between the three volumes we consid++er.
As we are focused on gas accretion from the cosmic web, we only consider smoothly accreted gas and exclude gas that is brought in by mergers from our analysis. However, in the low mass halo regime, smoothly accreted gas easily dominants the accreted gas budget \citep{Keres2005,Keres2009,L'Huillier2012,Chun2019}.

This paper is organized as follows. 
In Section \ref{sec:method}, we describe our zoom simulations and sample selection.
In \ref{sec:results}, we describe we uncover what affects the ability for a halo to maintain its gas against reionization, and what factors control the accretion of new gas from the cosmic web, and its impact on star formation histories. In Section \ref{sec:summary}, we summarize our results.

\section{Method}
\label{sec:method}

\subsection{Cosmological hydrodynamic simulations}
We perform three different cosmological hydrodynamic zoom simulations using the parallelized N-body/smoothed particle hydrodynamics (SPH) code, Gadget-3 \citep{Springel2005}.
The three different simulations are high resolution zooms on three different dwarf galaxies whose virial mass, $M_{vir}$, is $\sim10^{10}~h^{-1}M_{\odot}$ at $z=0$, respectively. They are zoomed from an 8~$h^{-1}$Mpc volume, in low density, field-like environment. We study all of the halos inside a zoom box of (1~$h^{-1}$Mpc)$^3$. 
To reproduce the properties of the galaxies and more small-scale gaseous structures, we include baryonic processes such as radiative cooling/heating, star formation, supernova feedback, cosmic reionization at $z\sim8.9$, and self-shielding for the dense gas with $n_H \geq$ 0.014 cm$^{-3}$.
In this paper, we describe the detail of the simulations briefly. 

Initial conditions for the simulations are generated at $z=49$ with the MUSIC package \citep{Hahn2011}, using the Planck cosmological model with $\Omega_{m} = 0.3,~\Omega_{\Lambda} = 0.7,~\Omega_{b} = 0.048$, and $h = 0.68$ \citep{Planck2014}.
17 million DM and 17 million gas particles are located in a (1~$h^{-1}$Mpc)$^3$ zoom box.
Outside of the zoom box, there are low resolution DM particles out to 8~$h^{-1}$Mpc in order to account for the effect of the surrounding cosmological structure.
The mass resolution of the high-resolution DM and gas particles is $\sim$ 4$\times$10$^3~h^{-1}$~M$_{\odot}$ and $\sim$ 8$\times$10$^2~h^{-1}$~M$_{\odot}/h$, respectively.
Radiative cooling/heating between 10~K and $10^8$~K is calculated using the  {\sc{CLOUDY90}} package \citep{Ferland1998}, including metal-line cooling.
Star formation occurs in cold dense gas clouds that satisfy: $n_H \geq$ 100 cm$^{-3}$, T $<$ 10$^4$ K, and $\nabla \cdot v <$ 0 \citep{Saitoh2008}.
The initial mass of a stellar particle is $\sim$ 270~M$_{\odot}/h$ when first formed, which is one third of the original gas particle mass. The star particles are assumed to consist of stars following the initial mass function of \cite{Kroupa2001}, and subsequently suffer stellar mass loss following the stellar evolution model of \cite{Hurley2000}.
For stars more massive than 8 $M_{\odot}$, we consider thermal energy feedback by supernova (SN) explosion \citep{Okamoto2008a}. The energy released by an explosion is fixed to $10^{51}$ erg, distributed amongst all gas particles within the SPH kernel radius.
The gravitational softening length in the zoom region is 130~$h^{-1}$~pc on a co-moving scale for $z>10$ and it is fixed as 12~$h^{-1}$~pc on a physical scale for $z\leq10$.
For more details, see \cite{Chun2019} and \cite{Shin2014}. 

We identify dark matter halos and their sub-halos using the Amiga Halo Finder (AHF), which groups gravitationally-bound particles, centered on a local density peak \citep{Knollmann2009}.
The virial radius, $r_{vir}$, of a halo is defined as the radius in which the average density is 200 times the cosmological background density at that redshift.

\subsection{Halo selection}
Cosmic reionization can ionise much of the gas within the cosmic web, and gas within dark matter halos that might otherwise become star-forming gas. This is especially important for low mass halos, whose small potential wells may struggle to hold onto their existing gas, and/or accrete new gas.
To study the impact of cosmic reionization, we select a sample of halos that form before cosmic reionization, and remain central halos (i.e., are not accreted into another halo) until after $z=6$. In this way, we can follow their evolution through reionization and beyond in order to see the effects on their gas content. We also demand that our sample has a gas mass that is more than 2.6$\times$10$^4~h^{-1}$~M$_{\odot}$ at the epoch of cosmic reionization, $z_{re}\sim8.9$. This choice ensures there are at least 32 gas particles initially, meaning we can follow the gas mass loss down to at least 3~$\%$ of the amount at reionization. In total, our sample consists of 381 halos in all three volumes.

\section{Results}
\label{sec:results}
\subsection{Time evolution of each halo's gas mass}

\begin{figure*}
\centering
\includegraphics[width=0.9\textwidth]{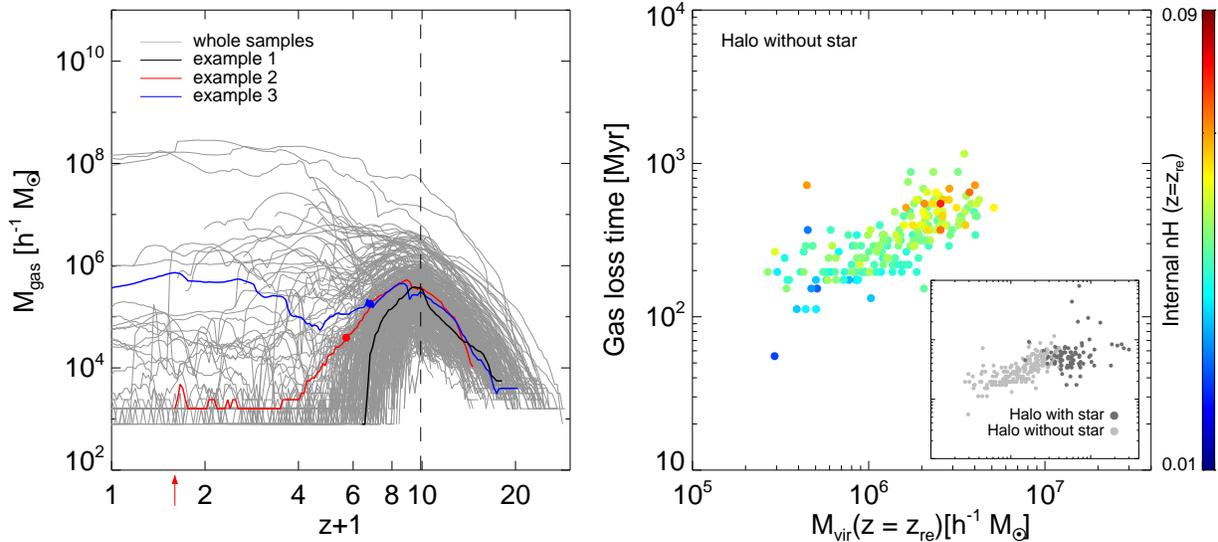}
\caption{The gas mass history of whole sample halos (left) and Scatter plot of gas loss time versus M$_{vir}$ (right). In left panel, black, red, and blue solid lines indicate the gas mass history of example 1, example 2, and example 3, respectively. The virial mass of the three examples is similar at $z_{re}$ (3$\times$10$^6~h^{-1}$~M$_{\odot}<~$M$_{vir}<~$3.6$\times$10$^6~h^{-1}$~M$_{\odot}$). The filled symbol indicates the redshift when the example re-accretes gas. The red arrow below the x-axis shows the redshift when example 2 falls into more massive halo. The vertical dashed line represents the epoch of cosmic reionization, $z_{re}\sim8.9$. In right panel, the color of filled symbols indicates the median value of hydrogen number density of gas in the halos at $z_{re}$. This scatter plot shows only the halos without stars. The small plot in the right panel shows the distribution of the halos with stars (dark gray) and without stars (light gray) in same x and y scales with the main plot.}
\label{fig:history}
\end{figure*}

The left panel of Fig. \ref{fig:history} shows the time evolution of the gas mass for our sample of halos. We can see that almost all of our sample (gray lines) accretes gas very efficiently before reionization (to the right of the vertical dashed line). In fact, in many cases this is the peak amount of gas that the halos will ever contain. Given that there dark matter masses are typically low at these early redshifts, this results in gas to dark matter fractions as high as 0.16. But very shortly after reionization, most halos quickly begin to lose their gas. We find that 38~\% of the full sample loses 90~\% of their gas content before $z=6$ (similar to the `example 1' case). 

In the right panel of Fig. \ref{fig:history}, we try to understand the reasons why some halos lose their gas very quickly while others hold onto their gas for much longer. We quantify the rate at which the gas is lost in terms of the gas loss time, shown on the y-axis. This is defined as the time interval between $z_{re}$ and time when a halo loses the gas mass below 10~\% of its gas mass at $z_{re}$. We see that the mass of a halo at reionization is a key parameter determining the gas loss time. Lower mass halos lose their gas more quickly. 

However, the variety of evolution in gas mass seen in the left panel of Figure \ref{fig:history} is not solely a function of halo mass. For example, example 1, 2 and 3 all have similar gas masses at $z_{re}$, and similar halos masses (3$\times$10$^6~h^{-1}$~M$_{\odot}<~$M$_{vir}<~$3.6$\times$10$^6~h^{-1}$~M$_{\odot}$), and yet show very different gas mass evolution following reionization. While they share a similar gas mass evolution until reionization, example 1 (black line) loses all of its gas content before $z=5$. Meanwhile example 2 (red line) loses its gas more slowly, with a small amount of re-accretion occurring later (filled circles indicate the start of re-accretion). Example 3 (blue line) re-accretes significant quantities of gas from $z\sim4$ onwards.

The vertical spread in the trend shown in the right panel of Fig. \ref{fig:history} indicates that halos of similar mass can have widely varying gas loss times. We find that the internal gas density of the halos at the time of reionization is an important factor in causing this spread. To calculate this quantity, we measure the median gas density of all particles that are bound to the halo. At fixed halo mass, gas mass loss times are shorter when the internal gas density is lower, and longer when the internal gas density is higher, creating a blue to red color gradient in the vertical direction across the vertical scatter in the trend. 
This can be understood physically in that halos with denser gas require more UV photons to be fully ionised. In fact, we confirmed that ionisation tends to first occurs in the less dense gas at large radius, and then proceeds radially inwards towards the galaxy centre.

We note that for the main panel on the right of Fig. \ref{fig:history}, we deliberately only show halos without any stars. This is in order to more clearly show the dependency on the internal density, without the scattering effect of stellar feedback which can both enhance gas loss times and reduce gas loss times (presumably by gas blow-out). The increased scatter is shown in the sub-panel comparing the halos with star particles (dark gray points), and without star particles (light gray points; same as colored points in main panel).
It can also be seen that the halos which form stars do not have exactly the same trend as is seen for halos without any stars. They appear mildly offset downwards, as if supernova feedback has slightly reduced their gas loss times by expelling some gas. Although we have only tested with one type of supernova feedback scheme, we can expect that if we had used strong feedback then perhaps the gas loss times would have been further reduced. The different choice of feedback scheme might also result in differing amounts of scatter about the trend.

\subsection{The parameter dependency on gas accretion}

\begin{figure*}
\centering
\includegraphics[width=0.75\textwidth]{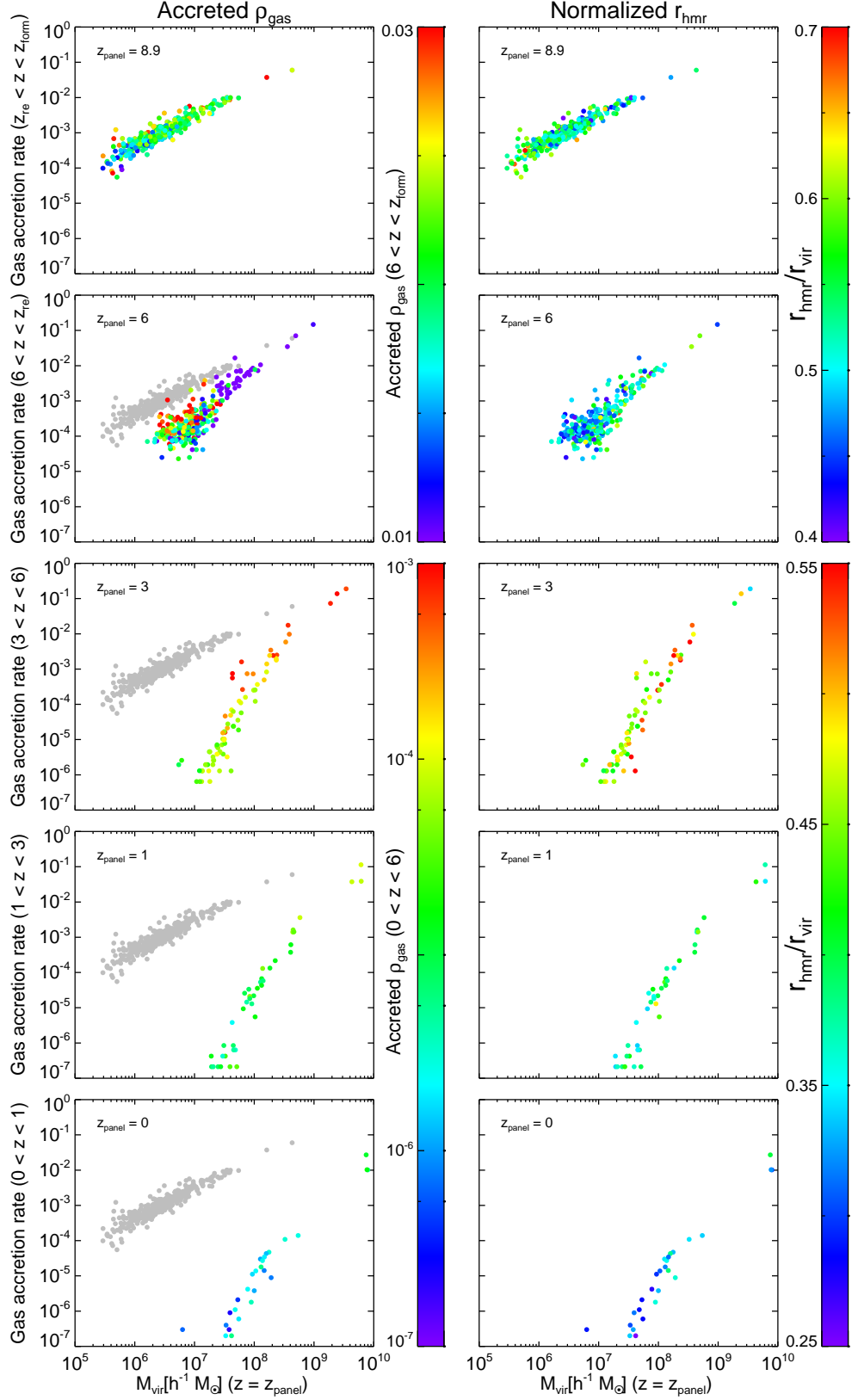}
\caption{Scatter plots of gas accretion rate versus M$_{vir}$ at $z_{panel}$. $z_{panel}$ is shown on the upper left edge of each panels. The dotted symbols are colored by the number density of gas that fall into the halo between each redshift bins (left) and the ratio of half mass radius to virial radius at $z_{panel}$ (right). The units of the gas accretion rate and the number density of gas are M$_{\odot}/h/yr$ and $cm^{-3}$, respectively. The two color scales next to each columns are shared for upper two panels and lower three panels, respectively. In the four lower left panels, the results at $z_{re}$ are plotted by gray colored symbols.}
\label{fig:accretion}
\end{figure*}

Clearly halo mass and internal gas density at reionization are key parameters for determining the gas mass loss after reionization. Next we try to what determines the internal gas density of the halos at reionization. We investigate two possibilities for having high internal gas densities; halos may accrete high density gas from the cosmic web (i.e. an external origin), or halos may efficiently compress and concentrate the lower density gas that they accrete (i.e. an internal origin).
For this, we first separate the whole sample of halos into the following five redshift bins: $0 \leq z < 1$, $1 \leq z < 3$, $3 \leq z < 6$, $6 \leq z < z_{re}$, and $z_{re} \leq z < z_{form}$.
Hereafter, we refer to the minimum redshift in each redshift bins as $z_{panel}$, as labelled in the upper left edge of each panel in Fig. \ref{fig:accretion}. For each halo, we measure the gas accretion rate, defined as the total accreted gas during that redshift interval divided by the time interval of the redshift bin.

\begin{figure}
\centering
\includegraphics[width=0.45\textwidth]{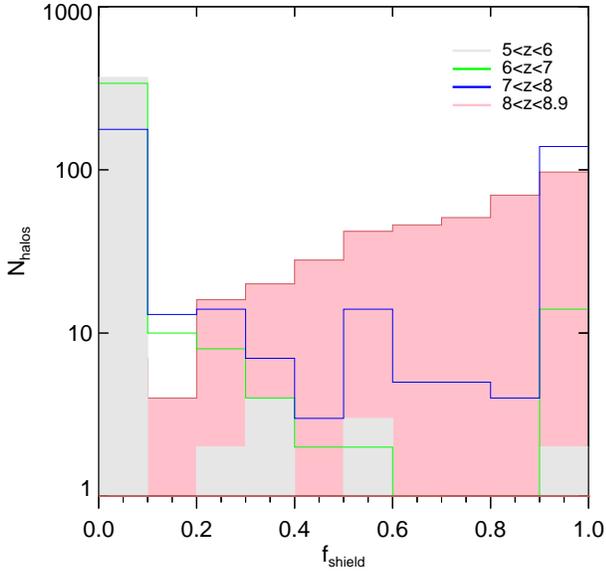}
\caption{A histogram of the fraction of the total accreted gas that is shielded, $f_{shield}$. Gray, green, blue, and red histograms indicate $f_{shield}$ during the various redshift ranges shown in the legend.}
\label{fig:shield}
\end{figure}

\begin{figure}
\centering
\includegraphics[width=0.45\textwidth]{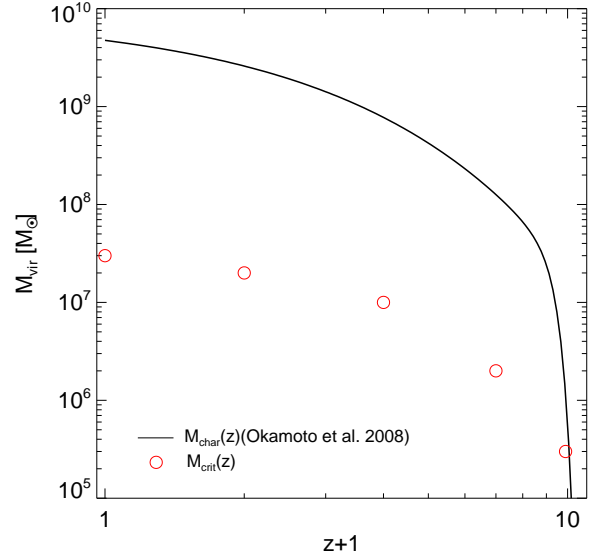}
\caption{The evolution of characteristic mass and critical mass. Solid black line represents the evolution of characteristic mass of \cite{Okamoto2008} and the red open circles indicate the critical mass, $M_{crit}$, at $z=0$, 1, 3, 6, and $z_{re}$.}
\label{fig:mc}
\end{figure}

Fig. \ref{fig:accretion} shows that the gas accretion rate of halos during all our redshift bin is strongly related to the virial mass at $z=z_{panel}$. The `Accreted $\rho_{gas}$' indicates the median value of the density of all gas that fall into the halo inside each redshift bin (left panels).
This parameter indicates whether the halos accrete high density gas or not.
To neglect the contribution of gas accreted by merging with other smaller halos, we only consider smoothly accreted gas in all of our plots.
The sudden change in the color-bar range after $z=6$ highlights how the typical density of externally accreted gas falls significantly after reionization. This is because reionization acts to heat up and lower the density of the cosmic web gas. In Figure \ref{fig:shield}, we plot a histogram of the fraction of externally accreted gas that is sufficiently dense to be self-shielding. We see that in the time period between $z=$8-8.9, and $z=$5-6 (approximately half a gigayear), the fraction of shielded accreted gas falls drastically to close to zero. This demonstrates that, even a lot of the cosmic web gas is self-shielding at reionization, within a short time period it can be evaporated away and replaced by hot, low density ionised gas, that is more difficult for dwarf galaxy halos to accrete. The lowest mass halos are the first to not be able to accrete gas anymore. We define a halo mass limit, $M_{crit}$, below which halos are unable to accrete gas. $M_{crit}$ increases from 3$\times$10$^5~h^{-1}$~M$_{\odot}$ to 3$\times$10$^7~h^{-1}$~M$_{\odot}$ between $z_{re}$ and $z=0$, as shown in Figure \ref{fig:mc}.

Figure \ref{fig:mc} also shows the characteristic mass defined in \cite{Okamoto2008}. This is defined as the halo mass when a halo has a baryon mass fraction of half the cosmic mean value in their simulations, and is meant to be a proxy for the mass below which halos can no longer accrete external gas at that redshift. As can be seen in the figure, our $M_{crit}$ values are always considerably smaller than their characteristic mass. In other words, we find that our low mass halos can continue to accrete gas, even when their mass is as small as 1~\% of the characteristic mass of \cite{Okamoto2008}. However, we caution that we are not comparing exactly the same quantities on the graph. We tried to calculate the characteristic mass in the same way as was done in \cite{Okamoto2008}. But, in our simulations, all of our halos have a baryon mass fraction less than half the cosmic mean value by $z=6$ due to gas loss by reionization. Therefore, it is not possible to directly compare between the simulations across the same range of redshift. Nevertheless, our low mass halos seem better able to accrete gas than was suggested in their study.

Some of these differences might, in part, stem from the higher halo mass and gas mass resolution of our simulations (see also Appendix of \cite{Chun2019} for further discussion).
\cite{Katz2019} investigate the inflow rate of the halos until $z=6$ using cosmological radiation hydrodynamic simulations. Although we cannot compare the gas accretion rate of our smallest halos with their results as we have higher resolution, we find that halos with $M_{vir}\sim10^7~h^{-1}$~M$_{\odot}$ in our two different works have similar gas accretion rates of $\sim2\times10^{-4}$~M$_{\odot}yr^{-1}$ at $z=6$. This supports the notion that even quite low mass dwarf galaxies ($M_{vir}\sim3\times10^7~h^{-1}$~M$_{\odot}$) can continue to accrete gas from the cosmic web at $z=0$, and at a far more efficient rate than was previously suspected. 

In all redshift bins we find that more massive halos accrete higher density gas (with the notable exception of $z_{panel}=6$ bin, to which we will return later).
We confirm that the density distribution of accreted gas and non-accreted gas from one to five virial radii from massive halos is very similar. Thus, it is not that more massive halos accrete high density gas selectively. Instead, it seems partly related to their location within the cosmic web, as we will show later in Section \ref{sec:environment}.

To test if, at fixed halo mass, some halos may be better at compressing and concentrating their accreted gas than others, we define the ratio $r_{hmr}/r_{vir}$, the ratio of half mass radius to virial radius of the dark matter (shown on the color bar of the right panels of Fig. \ref{fig:accretion}). This parameter essentially indicates how concentrated the halo is. We tried using the standard NFW halo concentration parameter but found the results were more scattered than when using $r_{hmr}/r_{vir}$ at fixed mass.
We can see that there is a hint of a trend for decreasing $r_{hmr}/r_{vir}$ with increasing halo mass in some redshift bins, but the main purpose of this plot is to look for a dependence on gas accretion rate at fixed halos mass (i.e. a vertical color gradient). We do see indications of some vertical color gradient. For example, red points tend to be slightly lower in the trend than green points in the $z_{panel}$=3 panel, meaning more concentrated halos tend to accrete more gas than less concentrated halos at fixed halo mass. 

However, the vertical color gradient seems a little clearer in the accreted gas density panels (left column) than in the right panel, especially at lower redshifts. We shall return to the issue of in which epoch the accreted gas density dominates and in which epoch the halo concentration dominates when we consider the next plot.

\begin{figure*}
\centering
\includegraphics[width=0.75\textwidth]{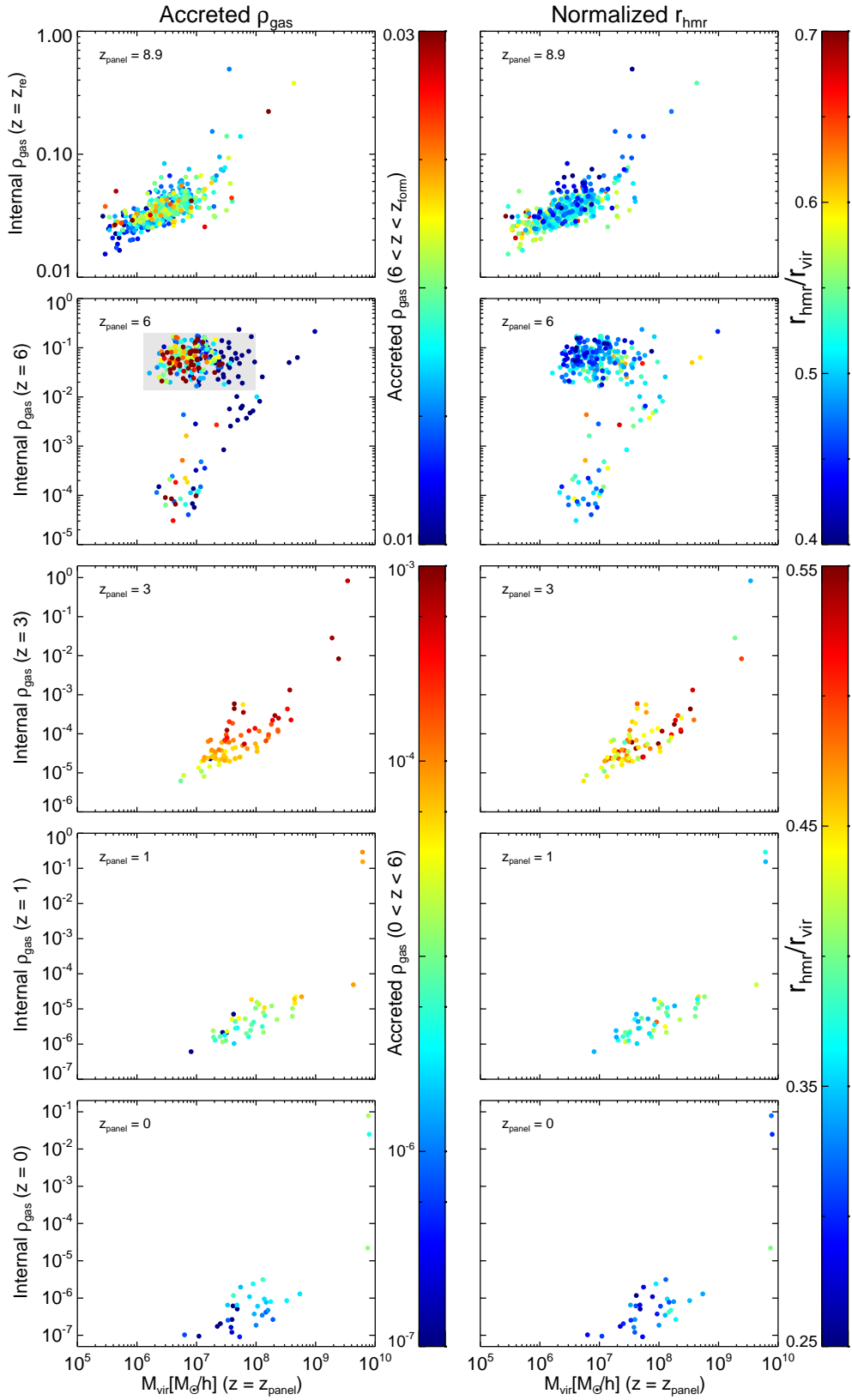}
\caption{Scatter plots of internal gas density versus M$_{vir}$ at $z_{panel}$. $z_{panel}$ and color scales in each panels are same with Fig. \ref{fig:accretion}. The shaded rectangular region in the left panel for $z_{panel}=6$ indicates the halos having high density gas even after $z_{re}$.}
\label{fig:nH}
\end{figure*}

As shown in Fig. \ref{fig:history}, we see that the internal density of the gas is an important factor in dictating how much gas loss occurs after reionization. Then in Fig. \ref{fig:accretion} we looked at the gas accretion rate. However, in principle a large amount of gas might be accreted but yet not be converted into dense internal gas, and remain in a low density state. So now, in Fig. \ref{fig:nH}, we consider how the internal gas density is related to the mass of the halo, and the accreted gas density (left column) and halo concentration (right panel).   
In all panels, we see a clear dependency on halo mass. Increasing halo mass results in higher internal gas density. We also see that, in general, higher halo masses tend to accrete denser external gas (i.e., the color of the points moves from bluer to redder colors). As we will show later in Section \ref{sec:environment}, this is in part a function of their location in the cosmic web. However, like in Fig. \ref{fig:accretion}, the $z_{panel}=6$ panel is the exception to the rule, with more massive halos appearing to accrete {\it{lower}} density gas. This occurs because only the massive halos have sufficiently large potential wells to be able to accrete hotter non-shielded gas. Meanwhile low mass halos can only accrete cooler, denser, self-shielded gas. In fact, the massive halos can also accrete this cooler denser gas, but because our measure of the accreted gas density is a median and the gas accreted by the massive halos is pre-dominantly hot, we tend to get lower values for the accreted density.

As the low mass halos are limited to only accreting cooler denser, self-shielded gas, and the amount of self-shielded gas is fast disappearing as it is evaporated away by reionization (Fig. \ref{fig:shield}), we find that $\sim$half of our sample of halos no longer accrete any gas after $z=7.7$, due to the surrounding gas becoming too hot due to reionization.
Therefore, we might expect that the gas mass growth history of a halo at recent times, long after reionization, might be more a function of the accreted gas density, rather than the halo concentration. Indeed, in the panels from $z$=3 onwards, we clearly see that it is the accreted gas density that affects the internal gas density more significantly than the concentration (i.e., there is a stronger vertical color gradient in the left column than in the right, in the lower three rows).

But at reionization (the top row), we can see a clearer dependency of the internal gas density on the halo concentration than on the accreted gas density. Thus prior to reionization, at fixed halo mass, the halo concentration seems to play an important role in compressing any accreted gas to a higher internal gas density, which in turn causes subsequent gas mass loss to occur at a slower rate (e.g. see Fig. \ref{fig:history}).

\begin{figure*}
\centering
\includegraphics[width=0.9\textwidth]{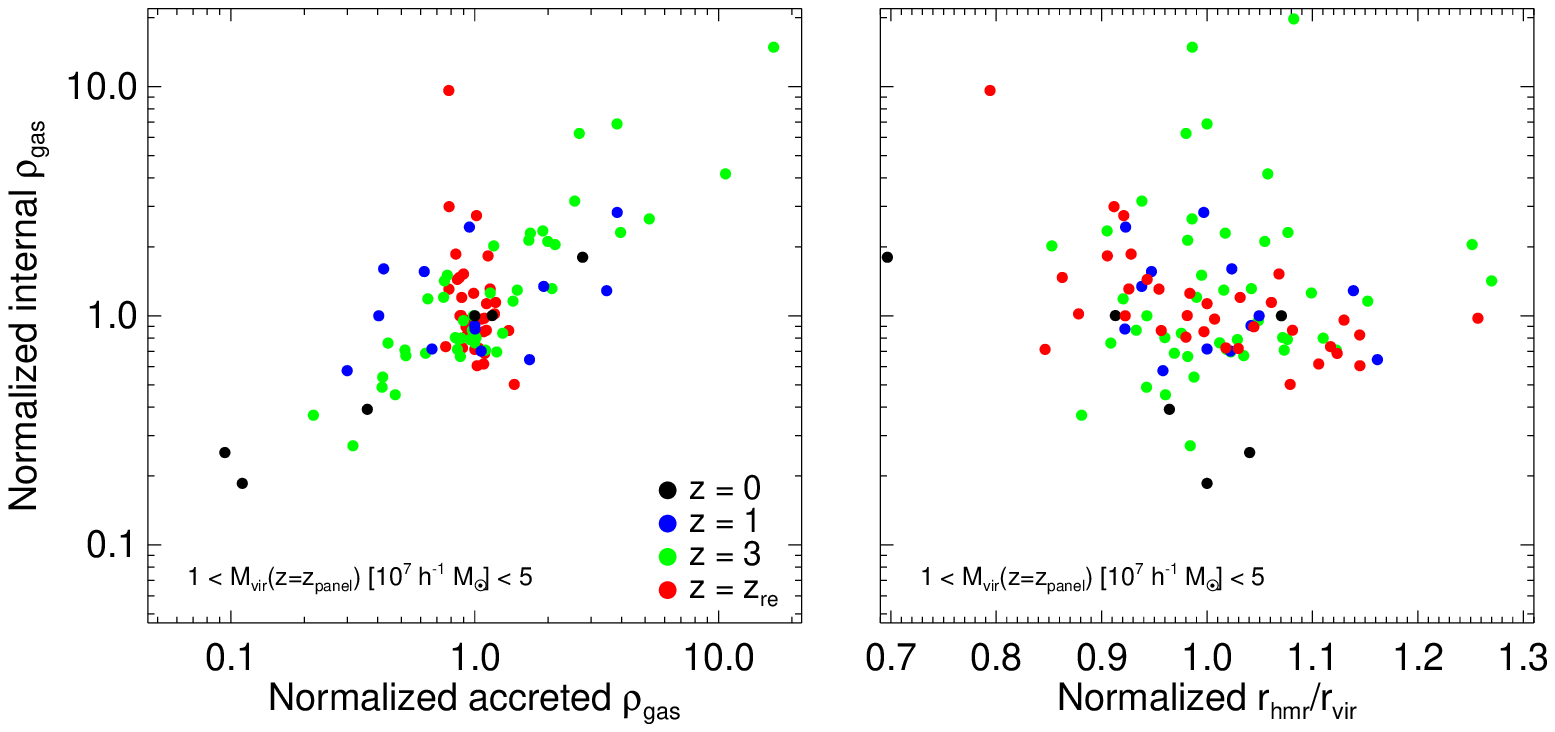}
\caption{Scatter plots of internal $\rho_{gas}$ versus accreted $\rho_{gas}$ (left panel) and $r_{hmr}/r_{rvir}$ (right panel) for halos within a narrow mass bin, as indicated in the lower left corner of the panels. Black, blue, green, and red filled symbols indicate the properties of the halos at $z_{panel}=0$, 1, 3, and $z_{re}$, respectively. Internal $\rho_{gas}$ and $r_{hmr}/r_{vir}$ are normalized by the median value of each properties at $z_{panel}$. Accreted $\rho_{gas}$ is normalized by the median values of the properties in each redshift bins.}
\label{fig:median}
\end{figure*}

To further confirm these relations in more detail, at fixed halo mass, we select halos in a narrow mass bin (with $M_{vir}$ of 10$^7~h^{-1}$~M$_{\odot}<~$M$_{vir}<~$5$\times$10$^7~h^{-1}$~M$_{\odot}$) in each redshift bin. For this, we exclude the results at $z_{panel}=6$ because the halos in the redshift bin are unusual compared to the halos at the other redshift bins, a fact we will discuss in more detail shortly in Section \ref{sec:starformation}. Fig. \ref{fig:median} shows the relation between internal gas density, accreted gas density, and concentration for all of the redshift bins. We normalise the internal gas density and $r_{hmr}/r_{vir}$ so that the trends can be easily compared by eye. To do this, we normalize by the median value of each property at each redshift (i.e., the median value for each redshift bin is found at the centre of the plot, at y=x=1).

As seen by eye previously, we confirm that there is a positive correlation between the internal gas density and accreted gas density can be seen for halos at low redshift ($z_{panel} \leq 3$). But the trend is not visible at $z_{re}$, and indeed the range of accreted gas density is very limited at this redshift. On the other hand, in the right panel we can see a negative correlation between the internal gas density and $r_{hmr}/r_{vir}$ parameter, which is not visible at more recent redshifts.

In summary, our results suggest that the density of the surrounding cosmic web gas that is accreted is important for determining the internal density of the halos at low redshift, but the concentration of the halos is more important at high redshift.

\subsection{Consequences for star formation}
\label{sec:starformation}

We note the presence of a cloud of points with high internal gas densities in panel $z_{panel}=6$, which we highlight with a shaded rectangular region. In fact, this cloud of high internal gas density points is the same as the trend seen in the panel above, $z_{panel}=8.9$. These are galaxies containing dense, self-shielded gas. After reionization, the amount of self-shielded gas available to be accreted drops rapidly (Fig. \ref{fig:shield}), and the internal self-shielded gas begins to be evaporated away. As a result, galaxies moves from the rectangular region down onto the much steeper trend, which is the same trend that can be seen in the more recent redshift panels below. The rate at which objects move from the rectangular region down to the steeper trend is also related to the evolving gas mass evolution histories shown in the left panel of Fig. \ref{fig:history}.

\begin{figure}
\centering
\includegraphics[width=0.45\textwidth]{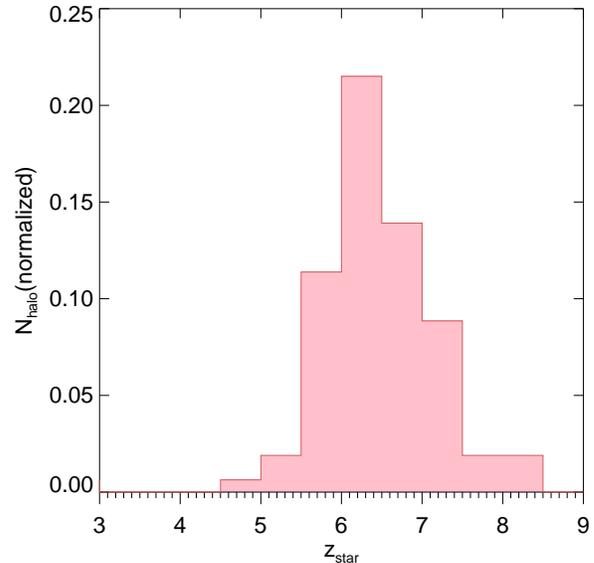}
\caption{The histogram of mean redshift of star formation in the selected halos. The number of the halos are normalized to 1.}
\label{fig:age}
\end{figure}

Prior to the loss of their dense gas, the halos in the shaded rectangular region have sufficiently dense gas to be able to form stars. We measure $z_{star}$, the mean redshift of star formation in each halo within the rectangular region, and show them in a histogram in Fig. \ref{fig:age}. Most of the star formation occurs near $z=6.0$ for these halos (54~\% of the halos have $z_{star}$ between $z=5.5$ and $z=6.5$). The star formation drops to nearly zero by $z=3$, as objects migrate from the rectangular region down to the steeper trend. We note that there was not much star formation at $z_{re}$ either, despite the fact that halos appear to have similar internal densities as at $z=6$. However, we are actually using median internal densities, and we confirm that at $z_{re}$ the peak densities of the gas are insufficient to result in star formation, as well as the gas being too hot.
In other words, even at $z=6$, only 0.35~\% of the total gas in halos is star forming, but at reionization this percentage was close to zero.

\subsection{Dependency on location in the cosmic web}
\label{sec:environment}

In Fig. \ref{fig:nH}, we suggested that the halos which accrete higher density gas are located in denser environments within the cosmic web. Thus, even at fixed halo masses, some small halos can accrete higher density gas than others at the same redshift. This is clearly demonstrated in the panel $z_{panel}=3$, where some low mass halos ($M_{vir}<~$10$^8~h^{-1}$~M$_{\odot}$) accrete gas as dense as that falling into the most massive halos. As we will now show, these small halos are in fact associated with the most massive halos, and accreting from the same reservoir of dense gas.

\begin{figure*}
\centering
\includegraphics[width=0.9\textwidth]{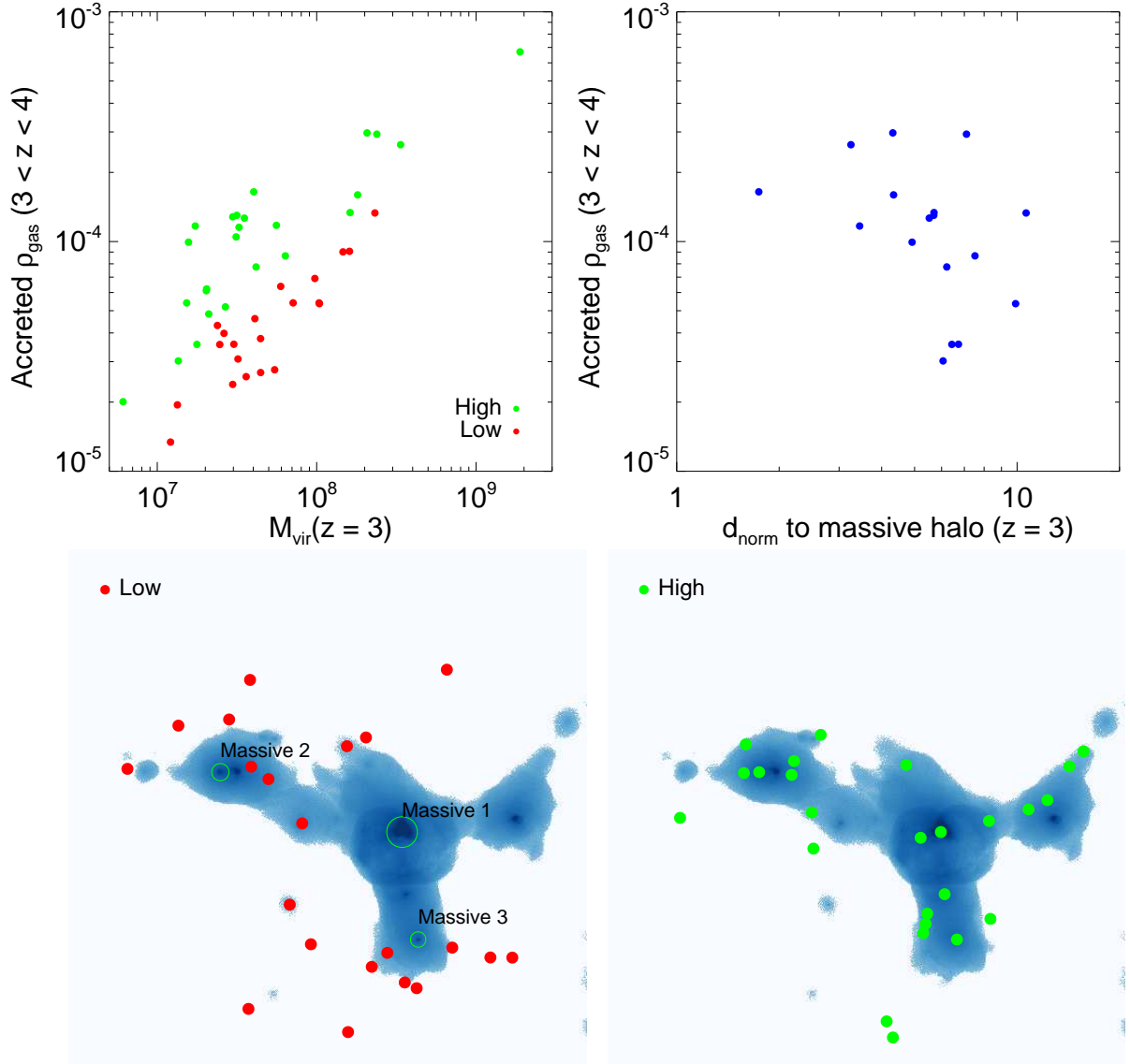}
\caption{The impact on density of accreted gas of halo location with respect to massive galaxies embedded in the cosmic web at $z_{panel}=3$. The upper left panel shows the median density of the gas that fall into the halos between $z=3-4$ versus $M_{vir}$. We split the sample into a high and low accreted gas density subsample, within a halo mass bin. In the lower left and lower right panels we show the location of the two subsamples at $z=3$, with respect to the dense cosmic web gas (n$_H$ $>3 \times 10^{-5}$~H/cc) that is concentrated on the three most massive halos and stretches along gaseous filaments linking the three massive halos. These massive halos are labelled in the left panel, and green circles show their halo virial radii. The volume shown is a cube of side-length 1~$h^{-1}$Mpc, centred on the most massive of the three massive halos, labelled `Massive 1'. The upper right panel shows the distance with respect to the nearest of the three massive halos at $z=3$, normalised by the virial radius of the massive halo.}
\label{fig:position}
\end{figure*}

To investigate this, we first consider the upper-left panel of Figure \ref{fig:position}. There is clearly a correlation between halo mass at $z=3$ and the median density of the externally accreted gas. However, we are interested in what drives the vertical scatter in this trend, therefore we split the sample into two subsamples, labelled as `High' and `Low' density accreters.

In the lower panels, we now consider the spatial distribution of these two subsamples, with respect to dense gas within the cosmic web (shown with a blue density plot). The dense gas is clearly concentrated about the three most massive halos $M_{vir}(z=3)~>~2\times10^8~h^{-1}$~M$_{\odot}$ within the view (shown with circles), and extends along filamentary structures that trace out the cosmic web in which the three massive halos are embedded.

By eye, it can clearly be seen that the `High' density accreters are located within regions filled with denser gas, and they clearly appear to trace out the cosmic web filamentary structure. Meanwhile the `Low' density accreters are much more scattered about, and located in lower density regions. Thus there appears to be a clear link between location in the cosmic web, and the density of the accreted gas.

It is already well known that more massive halos tend to be found closer to the cosmic web. But with our mass independent choice of the `High' and `Low' subsamples, we ensure that the dependence we see on location in the cosmic web is also valid at fixed halo mass.

Finally we consider the upper-right panel. Here we identify which of the three most massive halos is closest to a given halo, and measure the distance between them. We normalise the distance by the virial radius of that massive halo. We find a strong trend for the accreted gas density to increase with decreasing distance to the nearest of the three massive halos. This further confirms what could be seen by eye in the lower panels, that halos within the denser parts of the cosmic web accrete much denser gas, at least while they are beyond the virial radius of the nearest massive halo. At these distances, gas accretion can clearly be quite efficient along the cosmic web structure. However, once halos enter into the massive halos, gas accretion may halt and even be reversed by active ram pressure stripping, as reported in our recent paper \citep{Chun2019}.

In summary, we find that the density of accreted gas is an important secondary parameter, after halo mass, that can alter their internal gas densities. The accreted gas density is, in turn, a clear function of their location with respect to massive halos that are embedded within the cosmic web.

\begin{figure*}
\centering
\includegraphics[width=0.9\textwidth, trim = 0 0 0 0, clip]{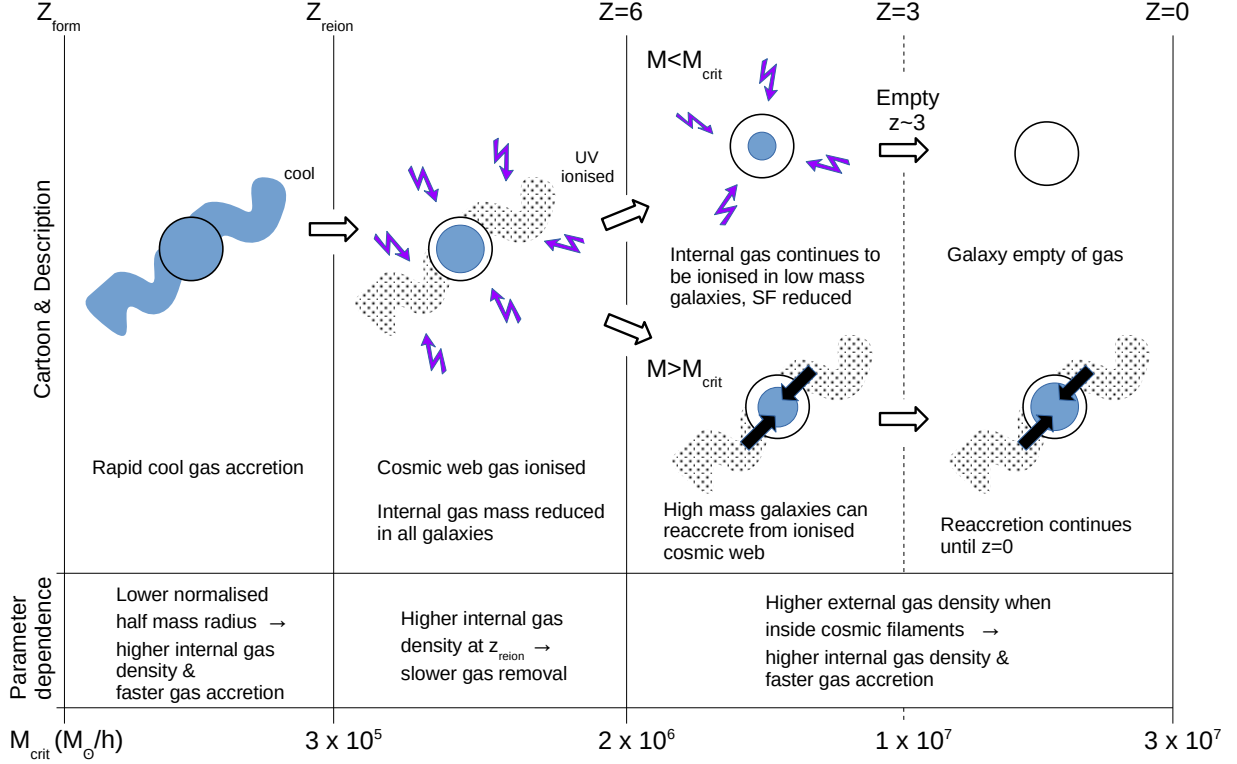}
\caption{Cartoon schematic of the gas accretion evolution of our sample of low mass halos. The evolutionary history is split into three main time periods (left to right); from formation time $z_{form}$ until reionization $z_{re}$, from reionization until $z=6$, and from $z=6$ until $z=0$.  Beneath a cartoon sketch and description, there is a row explaining the dependency on key parameters at that time.  In the bottom row, we give the critical mass $M_{crit}$ below which dark matter halos do not accrete any gas during that time period.}
\label{fig:cartoon}
\end{figure*}

\section{Summary and discussion}
\label{sec:summary}
In this work, we aim to investigate how low mass halos ($M_{vir}\sim3\times10^{5}~h^{-1}$~M$_{\odot}$ (at $z=z_{re}$) to $M_{vir}\sim10^{10}~h^{-1}$~M$_{\odot}$ (at $z=0$)) accrete cosmic web gas and respond to cosmic reionization. 

For this, we use three different high resolution cosmological simulations that 1$~h^{-1}$Mpc box zooms, selected from an 8$~h^{-1}$Mpc volume \citep{Chun2019}. In our previous paper, \cite{Chun2019}, we studied the fate of low mass halos after they fall in dwarf galaxy mass hosts. We found that tidal mass loss, combined with ram pressure stripping, is an efficient means by which accreted small halos can lose their gas content, even though the host galaxy is only a dwarf galaxy itself.
In this study, we wish to understand how low mass galaxies evolve their gas masses without the complicating influences of a host galaxy. Therefore, our sample consists of halos that are centrals at the epoch we are studying, and they simply disappear from our sample at the infall epoch, if they should become a satellite.

We summarize our main results on gas accretion histories of low mass halos, from their redshift of formation $z_{form}$ until $z=0$, in the cartoon schematic shown in Fig. \ref{fig:cartoon}. We describe each time period below:

1) {\it{$z_{form}$ to $z_{re}$}}: In this pre-reionization period, halos can accrete gas rapidly and efficiently (see left panel of Fig. \ref{fig:history}), and their fraction of gas to DM mass typically reaches to roughly the cosmic mean value, 0.16.
During this time, we find the internal density of the gas inside halos is mainly a function of their mass, but there is also a secondary dependence on the halo compactness, measured using the normalized half mass radius, $r_{hmr}/r_{vir}$. Here, a smaller value of $r_{hmr}/r_{vir}$ means a more compact halo, which is more effective at compressing externally accreted gas, and converts it into higher internal gas densities. At this time, we do not find a strong dependence on the density of gas that is accreted onto the halos.

2) {\it{$z_{re}$ to $z=6$}}: Following reionization, most of the cosmic web gas is ionized during a short time period by the UV radiation. This causes the lowest mass halos to be unable to accrete gas anymore. Below a halo mass of approximately $2\times10^{6}~h^{-1}$~M$_{\odot}$, virtually all halos lose their internal gas due to UV ionization during this time period. We find that the rate at which their internal gas is removed is primarily a function of their halo mass, with a secondary dependence on their internal gas density at the time of reionization. Physically this can be understood as the denser internal gas takes more time to be evaporated away, from the outside inwards. 

3) {\it{$z=6$ to $z=0$}}: After $z=6$, we find the main parameter controlling accreted gas accretion and the density of internal gas is halo mass, with a secondary dependence on the external environment. During this time period, the halo compactness plays a lesser role. Halos within the filamentary structure of the cosmic web, and especially those closer to (but not inside) massive halos in our sample ($M_{vir}~>~1.35\times10^{7}~h^{-1}$~M$_{\odot}$) tend to accrete denser gas than their equal mass counterparts in the field, and in turn this increases the internal density of gas inside of the halos. Distances of between roughly 1-5 virial radii from more massive halos may represent an ideal place to efficiently accrete dense gas, during this more recent time period (see the upper right panel of Fig. \ref{fig:position}).

Those halos with a lower mass than critical mass, $M_{crit}$, for gas accretion tend to lose their gas continuously over this time period, and thus their star formation is reduced and eventually halted. Eventually, these halos lose all of their gas content by $z=3$ and do not return to re-accrete fresh gas.
On the other hand, the more massive halos in our sample that have $M_{vir}~>~M_{crit}$ can re-accrete fresh external gas from the ionized cosmic web, and often continue to actively do so until $z=0$.

4) As the universe expands over time, the cosmic web gas density decreases and the gas temperature increases continuously.
This makes that the critical mass, $M_{crit}$, which is the minimum halo mass for gas accretion to steadily increase with time. We find that $M_{crit}$ increases from $M_{vir}\sim3\times10^{5}~h^{-1}$~M$_{\odot}$ to $M_{vir}\sim3\times10^{7}~h^{-1}$~M$_{\odot}$ between $z_{re}$ and $z=0$. These halo masses above which gas accretion begins are considerably smaller than the characteristic masses of \cite{Okamoto2008}. This means that even quite low mass field dwarfs ($M_{vir}(z=0)~>~3\times10^{7}~h^{-1}$~M$_{\odot}$) can continue to accrete fresh gas from the cosmic web at $z=0$, in a much more efficient manner than was previously thought possible.

Our results show that the density of the surrounding cosmic web gas that is accreted is important for determining the internal density of the gas in halos at low redshift, but the compactness of the halos is more important at high redshift.
As the internal density of the gas is directly connected to their star formation, these results indicate that the star formation of the halos is influenced by their location within the cosmic web, and this is strongest at more recent times, following reionization.

In this paper, we focus on the gas mass evolution of low mass halos that are found in low density, field-like regions.
However, in the future we will extend our study to investigate the gas evolution of low mass halos and their star formation in higher density regions such as in the vicinity of groups and clusters.

\acknowledgments
This work was supported by the National Research Foundation grant funded by the Ministry of Science and ICT of Korea (NRF-2014R1A2A1A11052367). This work was also supported by the BK21 plus program through the National Research Foundation (NRF) funded by the Ministry of Education of Korea.



\end{document}